\documentstyle[12pt,aasms4]{article}

\newcommand\wave[1]{\mbox{$\lambda$#1\,\AA}}

\newcommand\eps[1]{log~$\varepsilon$(#1)}
\def\kmsec{\mbox{km~s$^{\rm -1}$}}
\def\teff{\mbox{T$_{\rm eff}$}}

\def\BmV0{\mbox{(B-V)$^{\rm o}$}}
\def\VmK0{\mbox{(V-K)$^{\rm o}$}}
\def\MV0{\mbox{M$_{\rm V}^{\rm o}$}}
\def\MV{\mbox{M$_{\rm V}$}}

\def\etal{\mbox{{\it et al.}}}
\def\eg{\mbox{{\it e.g.}}}

\def\cs22892{{\rm CS~22892-052}}

\begin{document}

\title{Evidence of Multiple r-Process Sites in the Early Galaxy: \\
New Observations of CS~22892--052\altaffilmark{1}}

\author{
Christopher Sneden\altaffilmark{2}, 
John J. Cowan\altaffilmark{3},
Inese I. Ivans\altaffilmark{2},
George M. Fuller\altaffilmark{4}, \\
Scott Burles\altaffilmark{5},
Timothy C. Beers\altaffilmark{6},
James E. Lawler\altaffilmark{7} 
}

\begin{center}
To appear in {\it The Astrophysical Journal Letters}
\end{center}

\altaffiltext{1}{Based on observations obtained with the Keck~I
Telescope of the W. M. Keck Observatory, which is operated by the
California Association for Research In Astronomy (CARA, Inc) on behalf
of the University of California and the California Institute of Technology.}
  
\altaffiltext{2}{Department of Astronomy and McDonald Observatory,
University of Texas, Austin, TX 78712; chris@verdi.as.utexas.edu, 
iivans@astro.as.utexas.edu}

\altaffiltext{3}{Department of Physics and Astronomy,
University of Oklahoma, Norman, OK 73019; cowan@phyast.nhn.ou.edu}

\altaffiltext{4}{Department of Physics, University of California at 
San Diego, La Jolla, CA 92093-0319; gfuller@ucsd.edu}

\altaffiltext{5}{Department of Astronomy and Astrophysics,
University of Chicago, Chicago, IL 60637; burles@uchicago.edu}

\altaffiltext{6}{Department of Physics and Astronomy, Michigan State 
University, East Lansing, MI 48824; beers@pa.msu.edu}
 
\altaffiltext{7}{Department of Physics, University of Wisconsin, 
Madison, WI 53706; jelawler@facstaff.wisc.edu}

\begin{abstract}

First results are reported of a new abundance study of neutron-capture 
elements in the ultra-metal-poor (UMP; [Fe/H]~= --3.1) halo field giant 
star \cs22892.
Using new high resolution, high signal-to-noise spectra, abundances of 
more than 30 neutron-capture elements (Z~$>$~30) have been determined.
Six elements in the 40~$<$~Z~$<$~56 domain (Nb, Ru, Rh, Pd, 
Ag and Cd) have been detected for the first time in a UMP star. 
Abundances are also derived for three of the heaviest
stable elements (Os, Ir, and Pb).
A second transition of thorium, \ion{Th}{2} \wave{4086}, confirms
the abundance deduced from the standard \ion{Th}{2} \wave{4019} 
line, and an upper limit to the abundance of uranium is
established from the absence of the \ion{U}{2} \wave{3859} line.
As found in previous studies, the abundances of the heavier
(Z~$\ge$~56) stable neutron-capture elements in \cs22892\
match well the scaled solar system r-process abundance distribution.
From the observed Th abundance, an average age of $\simeq$ 16~$\pm$ 4~Gyr 
is derived for \cs22892, consistent with the lower age limit of 
$\simeq$~11~Gyr derived from the upper limit on the U abundance.
The concordance of scaled solar r-process and \cs22892\ abundances
breaks down for the lighter neutron-capture elements, supporting
previous suggestions that different r-process production sites
are responsible for lighter and heavier neutron-capture elements.

\end{abstract}


\section{Introduction}
Ultra-metal-poor (UMP) stars serve a critical role for understanding the 
initial epochs of our Galaxy; the observed abundances in these very old 
stars provide clues to the nucleosynthetic processes in the earliest
Galactic stellar generations. 
The UMP halo giant \cs22892\ ([Fe/H] = --3.1\footnote{
[A/B]~$\equiv$~log$_{\rm 10}$(N$_{\rm A}$/N$_{\rm B}$)$_{\rm star}$~--
log$_{\rm10}$(N$_{\rm A}$/N$_{\rm B}$)$_{\odot}$, and
\eps{A}~$\equiv$~log$_{\rm 10}$(N$_{\rm A}$/N$_{\rm H}$)~+~12.0,
for elements A and B. 
UMP stars are considered to be those with [Fe/H]~$<$~--2.5.}) 
merits special attention in nucleosynthesis studies. 
This star has extremely large overabundances of neutron-capture 
(n-capture) elements relative to iron, and the abundances of those 
elements with Z~$\ge$~56 apparently are consistent only 
with a scaled solar system r[apid]-process abundance distribution 
(Sneden \etal\ 1996; Cowan \etal\ 1999; 
Norris, Ryan, \& Beers 1997; Pfeiffer, Kratz \& Thielemann 1997).
In addition, this is the first star for which thorium and 
an extensive number of n-capture elements have been
detected, allowing an estimation of its radioactive age
(Sneden \etal\ 1996; Cowan \etal\ 1997, 1999; Pfeiffer \etal\ 1997).
Thus far, two n-capture element domains in \cs22892\ and other UMP
stars have been largely unexplored: the region 40~$<$~Z~$<$~56 
(between Zr and Ba); and the region 75~$<$~Z~$<$~83 
(the 3$^{\rm rd}$ n-capture peak peak, Os$\rightarrow$Pb).  
Using extensive new high resolution spectroscopic data, we have derived 
a new model-atmosphere for \cs22892, and derived abundances for a set 
of n-capture elements never before seen in this (or any other UMP) star.
These new abundances and their implications for early Galactic 
nucleosynthesis of n-capture elements are discussed in this {\it Letter}.

\section{Observations, Reductions, and Abundance Analysis}

Two new high resolution, high S/N spectra of \cs22892\ were obtained.
The Keck~I HIRES (Vogt \etal\ 1994) was used for one night in August, 1998 
to obtain a spectrum (the sum of multiple individual integrations) 
in the near-UV region,
3200~$\lesssim$~$\lambda$~$\lesssim$~4250~\AA.
The resolving power was R~$\equiv$~$\lambda/\Delta\lambda$~$\simeq$~45,000,
and the S/N varied smoothly from $\sim$30 near \wave{3200} to
$\gtrsim$200 near \wave{4250}.
We also used the McDonald 2.7m Smith telescope and 2d-coud\'e
echelle spectrograph (Tull \etal\ 1995) for three nights in
October, 1998 to obtain a high resolution 
(R~$\simeq$~60,000) spectrum, also the sum of multiple exposures,
at longer wavelengths (4400~$\lesssim$~$\lambda$~$\lesssim$~8000~\AA).
Again, the S/N varied with wavelength, from $\sim$50 near \wave{4500} to
$\gtrsim$150 near \wave{7000}.
Accompanying both HIRES and 2d-coud\'e observations of \cs22892\
were standard auxiliary integrations on tungsten filament and Th-Ar 
hollow cathode lamps.
For the 2d-coud\'e data, observations of hot, rapidly rotating
(essentially featureless) stars of similar airmass to \cs22892\
were obtained to facilitate removal of telluric spectral features
in the longer wavelength spectral regions.
Standard echelle reduction techniques were used to produce the final spectra. 

We determined new model atmosphere parameters for \cs22892\
from an analysis of equivalent widths (mostly from uncrowded longer 
wavelength spectral regions) of nearly 200 transitions of 
lighter (Z~$\leq$~30) elements.
This analysis applied the usual conditions that demand 
agreement between abundances from low- and high-excitation lines and 
from weak and strong lines of a given species, and agreement between 
neutral and ionized species abundances.
We interpolated model atmospheres from the Kurucz (1999) grid,
and employed the current version of Sneden's (1973) line analysis code.
The derived model has \teff~=~4710~K, log~g~=1.50,
v$_{\rm t}$~=~2.1~\kmsec, and model metallicity [M/H]~=~--3.2.
These values are in good agreement with the parameters
(4760~K, 1.30, 2.3~\kmsec, --3.1) determined previously by 
McWilliam \etal\ (1995) from lower resolution, lower S/N spectra.
Our new abundances for the Z~$\leq$~30 elements also agree well
with those of McWilliam \etal, but with greatly reduced 
line-to-line scatter.

With the new model atmosphere, abundances for n-capture elements
were determined from equivalent width and synthetic spectrum analyses.
The analysis techniques are described in detail by Sneden \etal\ (1996).
Table~1 gives the new abundances and their single-line 
standard deviation values, as well as abundances that were 
determined by Sneden \etal.
We first analyzed some n-capture elements (Y, Zr, Nd, Sm, Eu, Gd, 
Dy, Er, Tm, and Yb) previously treated by Sneden \etal.
These elements have large numbers of transitions in
the near-UV ($\lambda$~$<$ 4000~\AA).
The new analyses used the enlarged spectral coverage 
of our new \cs22892\ data and the most recent oscillator strength
information in greatly expanding the line lists for these elements. 
The reliability of the abundances increased, as 
indicated by lower $\sigma$ values for most of the re-analyzed elements.
The new abundances are in good accord with those of Sneden \etal. 
Moreover, there are no discernible abundance trends for individual 
n-capture element species with wavelength, suggesting that the
model atmosphere and line analysis techniques consistently reproduce the
observed \cs22892\ spectrum from the red to the near-UV
spectral regions.

We then searched the spectra for transitions of n-capture 
elements not previously detected in UMP (or in fact, in most) stars.
This yielded detections of six light n-capture elements
(Nb, Ru, Rh, Pd, Ag, and Cd).
Additionally, some elements of the 3$^{\rm rd}$ n-capture
peak (Os, Ir, and Pb) were identified; of this group only Os
had previously been tentatively detected by Sneden \etal\ (1996).
We derived abundances for these elements via synthetic spectrum
computations, since their transitions lie mainly in the crowded near-UV
spectral region.
For illustration, in Figure~1 we present the observed and synthetic 
spectra of lines for three of the elements.
These transitions and other new ones were easily identified even
in the lowest wavelength spectral regions.
In Table~1 we list the abundances of these newly identified elements.
Note that Nb, Ru, and Cd have only one transition so far identified on our 
spectra.
The (relatively large) $\sigma$ values for these elements represent
conservative estimates from assessment of transition probability 
information, continuum placement, identification of blending spectral
features, and synthetic spectrum fits.

Finally, we considered the radioactive chronometer elements
thorium and uranium.
We detected a new \ion{Th}{2} line at \wave{4086.52}, and analyzed that 
line in conjunction with the usually-employed \wave{4019.12} line.
The abundances from these two transitions are in excellent agreement,
and are very similar to the Th abundance determined by Sneden \etal\ (1996).
We did not find any features of \ion{U}{2}, but derived an
upper limit to the U abundance from an estimate of the
maximum strength of the un-detected \wave{3859.57} feature.

\section{Discussion and Conclusions}

In the top panel of Figure~2 we plot the \cs22892\ n-capture abundances 
from this study, and those of Sneden \etal\ (1996) for elements not 
analyzed by us.
A scaled solar system r-process elemental abundance distribution
is also shown.
The solar distribution is obtained by a decomposition of the solar system 
elemental abundances (Anders \& Grevesse 1989) into s[low]- and r-process 
contributions to individual isotopes, and is based upon the 
measured isotopic n-capture cross sections (K{\"a}ppeler \etal\ 1989,
Wisshak \etal\ 1996).
Summation of those contributions produces a
solar system r-process {\it elemental} abundance curve;
see Burris \etal\ (2000) for details of this procedure.
The solar system curve has been shifted to match the mean abundance level 
of the heavier n-capture elements (56~$\leq$~Z~$\leq$~72) in \cs22892.
The mean difference is 
$<$log~$\epsilon_\cs22892$~--~log~$\epsilon_{\rm s.s.}>$~=
--1.41~$\pm$~0.02 ($\sigma$~=~0.08, 15 elements).
The small scatter about the mean value confirms and extends all previous 
studies that have found consistency between \cs22892\ heavy n-capture element
abundances and the solar system r-process distribution.
In the bottom panel of Figure~2 this agreement is shown via a plot of 
the differences $\delta$(log~$\epsilon$) between 
the \cs22892\ abundances and the scaled solar curve.

Abundances of the 3$^{\it rd}$ n-capture peak elements Os, Ir, and Pb are 
also in good agreement with the scaled solar system r-process distribution. 
Abundances of 3$^{\it rd}$ peak elements in two other UMP stars
(\eg, Sneden \etal\ 1998) have anticipated this result, but these
are the first reliable 3$^{\it rd}$ peak abundances in \cs22892.
Thus the solar r-process pattern extends throughout the 
56~$\leq$~Z~$\leq$~82 element domain in this star.\footnote{
The Pb abundance is based on extremely weak transitions, and remains 
poorly determined.}
A nearly identical abundance distribution is observed in
the UMP star HD 115444 (Westin \etal\ 2000).
In addition, the [Ba/Eu] ratio  in most UMP stars
is in accord with the solar r-process 
value (\eg, McWilliam 1998, Burris \etal\ 2000). 
The agreement of all of these abundance patterns with the solar system 
r-process distribution suggests a uniform site, and/or uniform 
conditions for synthesis of the heavier n-capture elements. 

The thorium abundance of \cs22892\ lies below the solar r-process curve in 
Figure~2, indicating that radioactive decay of this element has taken place
over the time since it was created by the progenitor of this star.
We computed a Th-based radioactive age for \cs22892\
with various input assumptions (such as using both theoretical
r-process predictions and the observed solar system abundances).
The calculations are sensitive to small parameter changes and give a 
range of results with a average of $\simeq$~16~Gyr.
The error bars on the derived age, including both 
observational and theoretical uncertainties, are $\simeq$~4~Gyr.
An age estimate for \cs22892\ from the upper limit on the uranium abundance
can be done by comparing this limit with predictions (Cowan \etal\ 1999) 
for the long-lived uranium $^{238}$U isotope ({\it i.e.}, assuming that we 
are not observing any of the relatively quickly decaying $^{235}$U isotope).
This yields a lower age limit of $\simeq$ 11~Gyr for \cs22892.
While this limit is even more uncertain than values based upon the 
detection of Th, it does provide a lower bound on the age of 
this star which, within the error uncertainty, is consistent with 
the age determination using the Th chronometer.

In contrast to the heavy stable elements, our observations clearly demonstrate
that the agreement between \cs22892\ and
solar system r-process abundances fails for the lighter (Z~$<$~56) 
n-capture elements (see Figure~2).
For these nine elements, 
$<$log~$\epsilon_\cs22892$~--~log~$\epsilon_{\rm s.s.}>$~=
--1.72~$\pm$~0.07 ($\sigma$~=~0.20). 
Abundances of six of the lighter n-capture elements lie well below the solar 
system r-process curve that reproduces the heavier elements. 
The abundances of the odd-Z light n-capture elements are substantially less
than those of the even-Z elements, a pattern typically seen in s-process 
nucleosynthesis. 
However, the scaled solar system s-process abundance distribution is
a poor match to the \cs22892\ lighter n-capture abundances.
Numerical experiments, analogous to those conducted by Cowan \etal\ (1995),
suggest that a mix consisting of solar r-process abundances plus 10\% of 
the solar s-process abundances can roughly fit the \cs22892\ data.
But the Y and Ag abundances of \cs22892\ are 0.3-0.5~dex lower than
this hybrid solar system distribution.
In addition, the overall abundance level of these elements is still
about 0.2~dex below the level of the heavier elements, when compared
together to the solar distribution (Figure~2).
There does not appear to be a simple way to mix solar n-capture abundances
to match those of \cs22892.

The existence of two distinct r-process signatures in solar system 
meteoritic material, one for n-capture nuclei lighter than mass number 
140 and one for heavier nuclei, has been previously suggested by 
Wasserburg, Busso, \& Gallino (1996),  Qian, Vogel, \& Wasserburg (1998).
The clear differences between the abundances of the heavier and the 
lighter r-process elemental abundances in CS 22892--052
are consistent with that suggestion.
Thus lighter and heavier nuclei possibly could be produced on different 
Galactic timescales and come from supernovae of different mass ranges
(Qian \& Wasserburg 2000).
Alternatively, neutron-star binaries also could be a source for 
one of the mass ranges of r-process nuclei (Rosswog \etal\ 1999). 

The total n-capture \cs22892\ abundance pattern is also consistent with 
a neutrino-heated supernova ejecta r-process in a single supernova event, 
albeit with two different epochs in the explosion/ejection process 
({\it cf.}, Woosley \etal\ 1994, and references therein). 
In neutrino-heated ejecta nucleosynthesis models the abundance yields are
extremely sensitive to the electron fraction in the shock re-heating epoch 
when the lighter r-process species are synthesized (Hoffman \etal\ 1996). 
The electron fraction and the resulting lighter r-process abundances
in this early epoch are expected to be different for each supernova event. 
By contrast, the later neutrino-driven wind epoch, where the heavier 
r-process nuclides originate in these models, should have similar 
conditions, therefore producing the same abundance pattern, in all supernovae.

An additional clue about early Galactic n-capture nucleosynthesis
lies in the now well-documented very large star-to-star scatter in 
the bulk [n-capture/Fe] ratios of UMP stars (\eg, McWilliam \etal\ 1995,
Burris \etal\ 2000).
In \cs22892\ for example, [Eu/Fe]~$\sim$~+1.6 (Sneden \etal\ 1996), 
while other UMP stars have [Eu/Fe]~$<$~0.
The large scatter in overall n-capture element content
is an indication of the chemical inhomogeneity of the Galactic halo;
the Galaxy was not well-mixed at very early epochs.
Possible explanations for the early scatter in the [n-capture/Fe] ratio 
involve separating the iron and r-process production into different 
types of supernovae with an initial iron production from very massive stars 
(Wasserburg \& Qian 2000).
Detailed abundance analyses of many very low metallicity stars, and more 
extensive theoretical r-process calculations will be needed to understand 
better the differences between the production of iron and the entire
range of r-process elements in the earliest Galactic stellar populations.

\acknowledgments

We thank Andy McWilliam, George Preston, Jim Truran, and Craig Wheeler
for helpful discussions, and the referee, Jerry Wasserburg, for 
helping us to improve the paper.
We are grateful to the staff of the Keck Observatory for their expert
assistance with the observations. 
This research was funded in part by NSF grants 
AST-9618364 to CS, AST-9618332 to JJC, AST-9529454 to TCB, and
PHY-9800980 to GMF.

\clearpage

\begin{table}
\dummytable\label{table1}
\end{table}

\clearpage

\begin{center}
{\bf Figure Captions}
\end{center}
 
\figcaption{
Observed spectra (filled circles) and synthetic spectra (solid lines)
are shown for three representative transitions of newly detected n-capture
elements in \cs22892.
The abundances assumed in generating the synthetic spectra of \ion{Pd}{1} 
$\lambda$3404 \AA (top panel), \ion{Ag}{1} $\lambda$3281 \AA (middle
panel), and \ion{Ir}{1} $\lambda$3514 \AA\ (lower panel) are noted
in the respective panels.
\label{fig1}}

\figcaption{In the top panel, n-capture abundances in \cs22892 are plotted 
as filled circles with error bars, along with a scaled solar system 
$r$-process abundance curve that is plotted as a solid line.  
See the text for explanation of how the match between stellar and solar 
system distributions was accomplished. 
Note also the abundance of Th, sitting well below the scaled solar 
system curve, and the (very low) upper limit for U indicated by an 
open symbol with an arrow.
In the bottom panel, a differential comparison is presented between 
individual elements and the scaled solar system r-process 
abundance distribution.
The ordinate $\delta$(log~$\epsilon$) represents the difference between the
abundance of an individual element and the scaled solar curve.
Again, an open symbol with an arrow denotes the U abundance upper limit.
\label{fig2}}

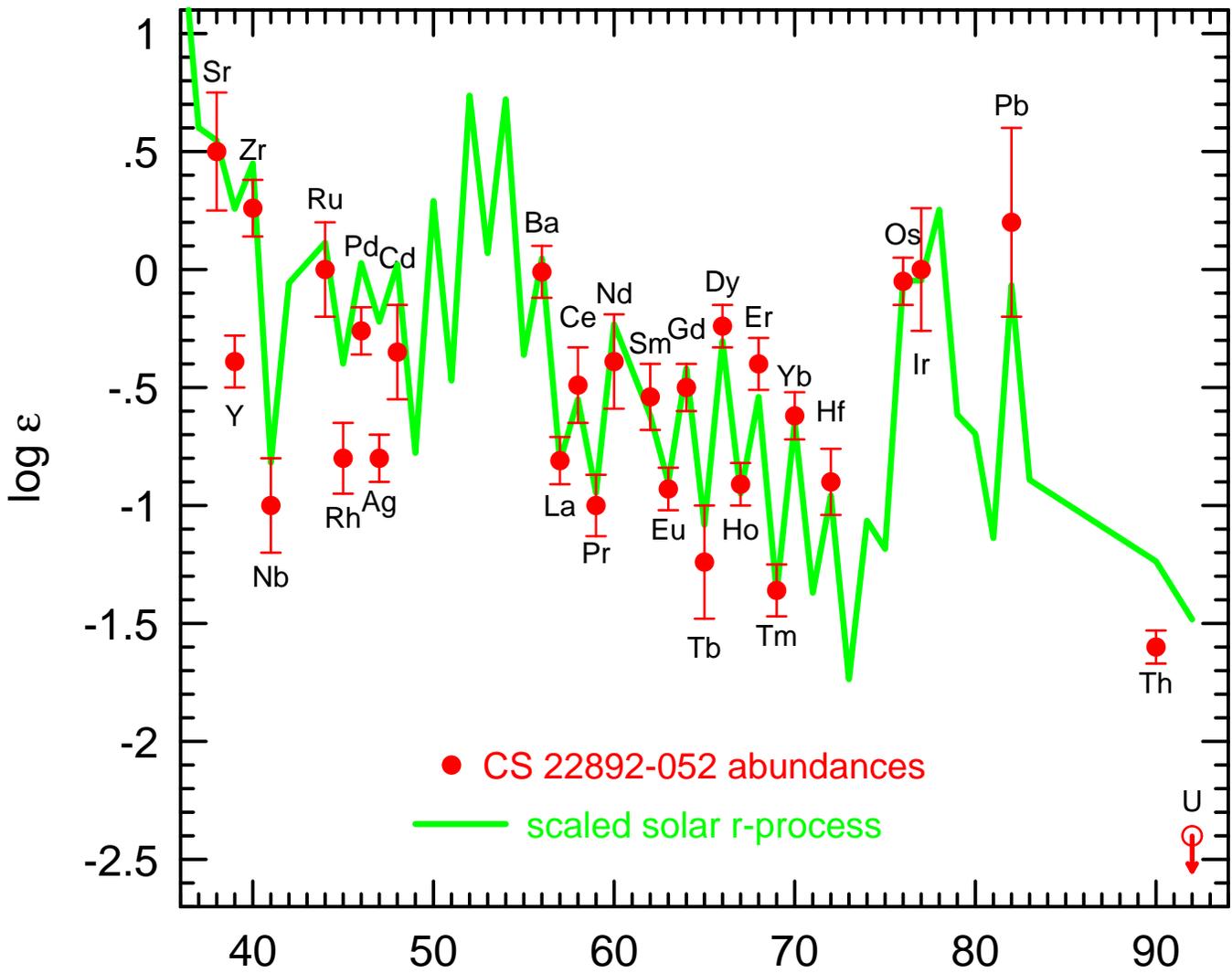
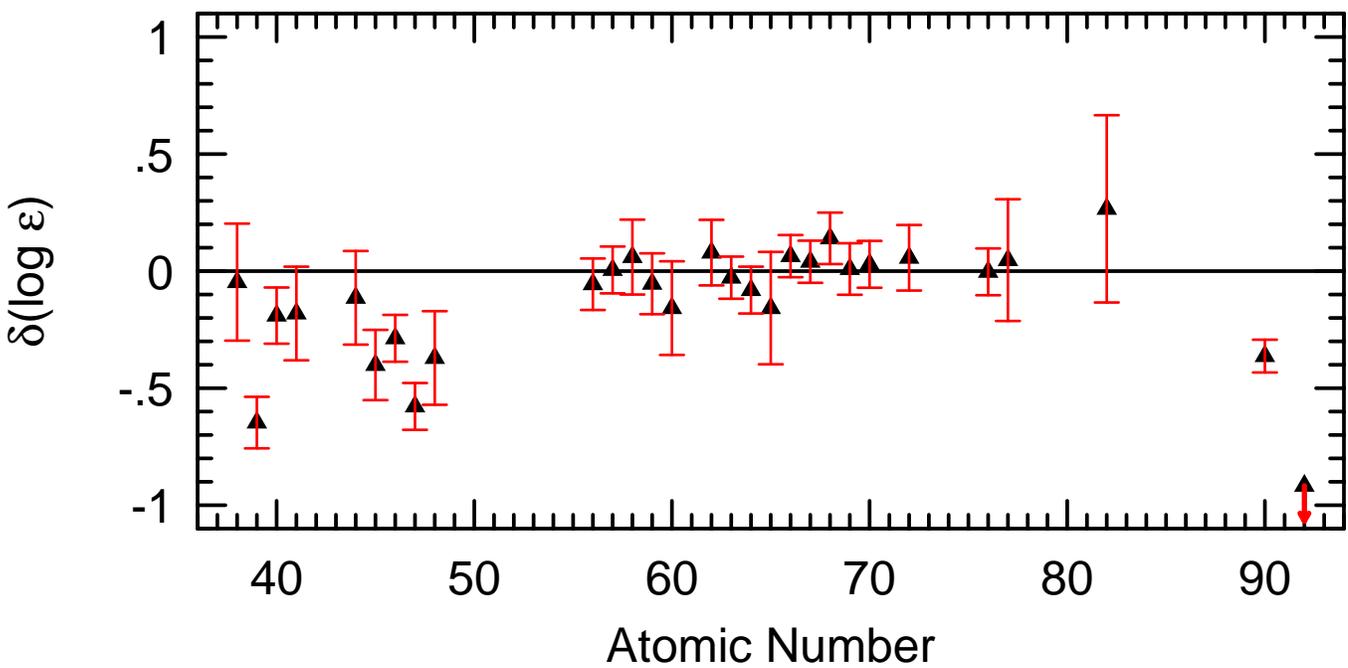
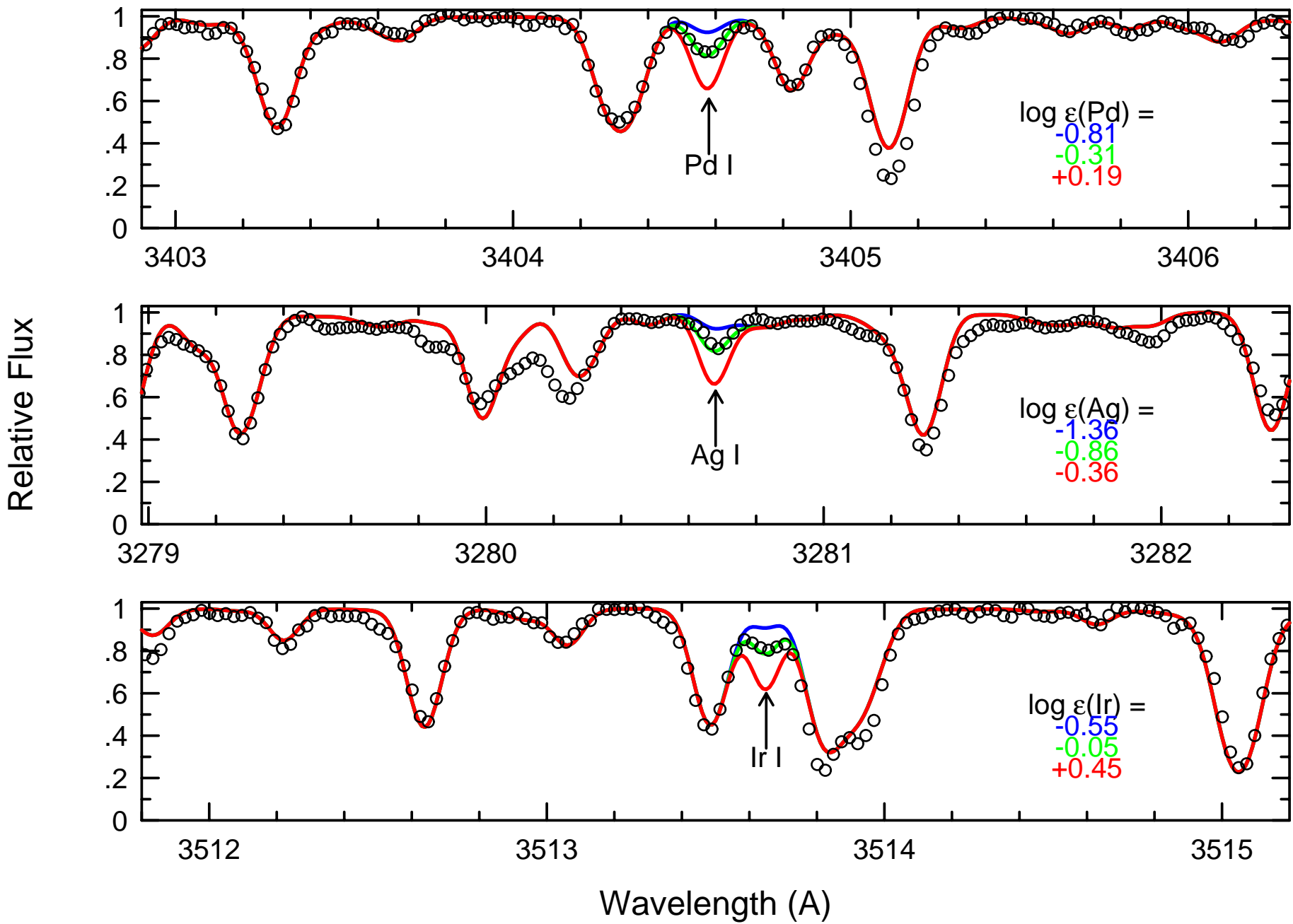
\end{document}